\documentclass[12pt]{article}\usepackage[hyperfootnotes=false]{hyperref}
\usepackage{epsfig}
\usepackage{float}
\usepackage{empheq}
\usepackage{bbold}

\usepackage[utf8]{inputenc}
\usepackage{amsmath}

\usepackage{caption}

\usepackage{amsmath}
\usepackage{amssymb}
\usepackage{graphicx}
\newlength{\abstractwidth}
\renewcommand{\thefootnote}{\fnsymbol{footnote}}
\renewcommand{\thanks}[1]{\footnote{#1}} 
\newcommand{\starttext}{
\setcounter{footnote}{0}
\renewcommand{\thefootnote}{\arabic{footnote}}}

\newcommand{\be}{\begin{equation}}
\newcommand{\bea}{\begin{eqnarray}}
\newcommand{\eea}{\end{eqnarray}}
\newcommand{\beq}{\begin{equation}}
\newcommand{\ee}{\end{equation}}

\newcommand*\widefbox[1]{\fbox{\hspace{2em}#1\hspace{2em}}}

\def\eq{&=&}

\def\la{\langle}
\def\ra{\rangle}
\def\simleq{\; \raise0.3ex\hbox{$<$\kern-0.75em
\raise-1.1ex\hbox{$\sim$}}\; }
\def\simgeq{\; \raise0.3ex\hbox{$>$\kern-0.75em
\raise-1.1ex\hbox{$\sim$}}\; }

\def\bi{\begin{itemize}}
\def\ei{\end{itemize}}

\def\sc{\setcounter{equation}{0}}

\def\CC{{\cal{C}}}

\def\CJ{{\cal{J}}}

\def\CL{{\cal{L}}}

\def\Tr{\rm Tr \it}
\def\bsub{ \begin{subequations}
\begin{empheq}[box=\widefbox]{align}  }
\def\esub{ \end{empheq}
\end{subequations}}

\def\1{\(  \mathbb{1} \)}

  \def\kl{k-local}

 \def\lf{\left(}
    \def\rg{\right)}

  \def\bn{\bigskip \noindent}

 \def\tb{\tilde{\beta}}
 \def\bm{\begin{bmatrix}}
 \def\em{\end{bmatrix}}
 \def\P{{\pm}}
  \def\M{{\mp}}
  \def\dr{\dot{\rho}}
  \def\ap{\approx}
  \def\msyk{Majorana-SYK}
  \def\nb{{\bar{\nu}}}

\makeatletter
\g@addto@macro\normalsize{%
  \setlength\abovedisplayskip{10pt}
  \setlength\belowdisplayskip{20pt}
  \setlength\abovedisplayshortskip{10pt}
  \setlength\belowdisplayshortskip{20pt}
}
\makeatother

\usepackage{color}

\begin{document}


\begin{titlepage}

\rightline{}
\bigskip
\bigskip\bigskip\bigskip\bigskip
\bigskip

\centerline{\Large \bf { Electric Forces in the  Charged SYK Model }}
\bn

\bigskip
\begin{center}
\bf      Leonard Susskind  \rm

\bigskip
 Stanford Institute for Theoretical Physics and Department of Physics, \\
Stanford University,
Stanford, CA 94305-4060, USA \\

and

Sandbox@Alphabet, \\
Mountain View, CA

\end{center}

\bn

\begin{abstract}
The connection between gravitational force and operator growth, reported in earlier papers, is generalized to include electromagnetic forces. It is shown how in the $U(1)$-invariant SYK system electric forces emerge through the same mechanisms---the momentum-size correspondence, and operator growth--- that give rise to gravitational force. The unit of electric charge implied by the $U(1)$-SYK theory is consistent with a Kaluza-Klein  radius inverse to the characteristic energy scale of the  SYK theory.


\end{abstract}

\end{titlepage}


\starttext \baselineskip=17.63pt \setcounter{footnote}{0}


\tableofcontents


\sc
\section{Introduction}\label{sec: Introduction}

The size of an operator \cite{Roberts:2018mnp} is a measure of the average number of elementary operators that are present in its expansion. For times shorter than the scrambling time\footnote{We will assume that the Hamiltonian is of the fast-scrambling type such as all-to-all \kl \ Hamiltonians  or sparse  Hamiltonians  of the type discussed in \cite{Xu:2020shn}} it is also a measure of the number of elementary operations that are required to prepare the operator, i.e., operator complexity. During the early phases of evolution,  complexity and size are indistinguishable. This paper is about that early phase; therefore I will use the symbol 
$\CC$ 
 for both.

While the concepts of size and complexity have not been precisely formulated for a continuum quantum field theory, they are well-defined for systems composed of a finite number of qubits, or as in the case of the SYK model, a finite number of fermionic operators.

The momentum-size correspondence  \cite{Susskind:2018tei}\cite{Brown:2018kvn}\cite{Susskind:2019ddc} is a holographic quantum-mechanical relation between the size of an operator and the radial momentum $P$ of the system created by acting with that operator. Roughly speaking, it says that size and momentum are proportional to one another, with a dimensional proportionality factor $\tb$ which may depend on the radial location of the particle. It is also possible to think of $\tb$ as a function of the time by following it along a light-like trajectory. We will return to the meaning of $\tb$. 

In the context of the SYK model an alternative formulation of the momentum-size correspondence was given in  \cite{Lin:2019qwu}\cite{Susskind:2019ddc},
\be
P =\frac{d\CC}{du}
\label{P=dC/du}
\ee
where $u$ is the ordinary time conjugate to the SYK Hamiltonian. 

Operator size is not conserved and in many contexts one sees a tendency for it to increase with time \cite{Roberts:2018mnp}, a phenomenon called operator growth\footnote{Because quantum mechanical evolution is reversible operators can also shrink, but we will include this possibility under the general heading of operator growth.}.
Since the force on a system is  the time-rate-of-change of its momentum, the momentum-size correspondence
implies a relationship between force and operator growth, and therefore between dynamics and quantum information. This has been illustrated  for gravitational force, but it should also apply to other forces such as electromagnetic.

Recent studies of these connections \cite{Susskind:2018tei}\cite{Brown:2018kvn}\cite{Susskind:2019ddc}
have concentrated on the  gravitational force of attraction on a particle falling into the long  AdS(2)  throat of a near-extremal SYK black hole. Although the SYK black hole is in many respects very similar to a near-extremal Reissner-Nordstrom  (NERN) black hole, the  usual Majorana-SYK system does not carry a conserved $U(1)$ charge, and therefore does not provide an opportunity to explore the mechanism leading to electric  forces. However the model can be generalized to have a global $U(1)$ symmetry \cite{Davison:2016ngz} by replacing the real fermionic degrees of freedom by complex fermions, $\psi$ and $\psi^{\dag} $  \cite{Davison:2016ngz}\cite{Gu:2019jub}. The model will be $U(1)$ invariant if each term in the Hamiltonian has an equal number of  $\psi$s and $\psi^{\dag}$s.

Standard lore says that if a holographic theory has a global symmetry then the bulk system that it describes will have a gauge symmetry. In the case of the $U(1)$-SYK system the bulk dual is the long throat of a near extremal black hole. As such it is a $(1+1)$-dimensional system with no propagating gauge degrees of freedom, but there will be one-dimensional  Coulomb forces  between charges. In particular a charged black hole should exert forces on charged particles in the throat---attractive for  opposite-sign  and repulsive for like-sign charges. If we combine this observation with the momentum-size correspondence it predicts  a difference in the growth rates for positively and negatively charged fermionic operators. The purpose of this paper is to see the ``nuts and bolts" of how this occurs in the quantum mechanics of the $U(1)$-SYK system.

In \cite{Brown:2018kvn}\cite{Susskind:2019ddc} two calculations were compared. The first was a general relativity calculation of the proper spatial momentum $P(t)$ of a neutral particle falling into a NERN background. The second was a microscopic calculation \cite{Qi:2018bje}
of  the  size of the Heisenberg operator  $\psi(t)$ in standard SYK which made no use of any assumption of a gravitational dual. Using  the momentum-size correspondence the calculations were shown to agree in detail.

In this paper we generalize this strategy to the $U(1)$-SYK model by  calculating the motion of a charge $\pm 1$ particle falling through the long throat of a near extremal $U(1)$-charged black hole. The acceleration (rate of change of momentum) of the particle depends on whether the particle and black hole have the same or opposite sign charges.
We then study the operator growth of the charged operators $\psi(t)$ and $\psi^{\dag}(t)$. For a given  charge on the black hole we find that the operator growths of $\psi(t)$ and $\psi^{\dag}(t)$ differ in just the way predicted by the momentum-size correspondence.

 \sc
\section{Preliminaries}

\subsection{The Bulk}

In previous papers   \cite{Brown:2018kvn}\cite{Susskind:2019ddc}  the Majorana-SYK model was compared with the dynamics of neutral particles in a background  of a dimensionally reduced Reissner-Nordstrom black hole. In this paper we will replace the Reissner-Nordstrom  background with the closely related but slightly simpler solution of the $(1+1)$-dimensional dilaton-gravity theory discussed in 	\cite{Maldacena:2016upp}. The model is a variant of the Jackiw-Teitelboim (JT) theory described by the action,
\be 
I =  -\frac{1}{16\pi G} \left[
\int d^2x \sqrt{g }  \lf  \Phi_0 +\phi   \rg R  + \frac{1    }{\mu^2}  \sqrt{g} \phi   +\rm BT + MT \it
\right]
\label{JT}
\ee
where  $\Phi_0$ is a dimensionless constant, $\phi$ is a dynamical field, BT stands for a boundary term, and MT stands for matter terms. The term proportional to $\Phi_0$ is topological and determines the zero-temperature extremal entropy. We will define the dilaton field $\Phi$ by,
\be
\Phi(x) = \Phi_0 + \phi(x).
\label{phi2}
\ee
The parameter $\mu$ has units of length and determines the radius of curvature of the  $AdS(2)$ solution.

The variational equation  for $\phi$ requires the curvature to be constant,
\be 
R=-1/\mu^2.
\label{R=-1/mumu}
\ee
and the equations generated by variation with respect to the metric are,
\be 
\nabla_{\mu} \nabla_{\nu} \phi  -g_{\mu \nu} \nabla^2 \phi +g_{\mu \nu} \phi =0
\label{phieq}
\ee
which, together with boundary conditions, determine $\phi.$

The solution of the equations of motion for the metric is $AdS(2)$ which can be written in the form,
\be 
ds^2 =  -\frac{r^2 -\mu^2}{\mu^2} dt^2 + \frac{\mu^2}{r^2-\mu^2} dr^2.
\label{metric}
\ee
The point $r=\mu$ represents the horizon but
unlike in higher dimensions the coordinate $r$ does not signify the radius of a sphere. Nor is its value at the horizon related to entropy. In JT-gravity that role is played by the dilaton field $\Phi(r)$ whose value at the horizon determines the  entropy according to,
\be 
S=\frac{\Phi(\mu)}{4G}.
\label{S=phi/4G}
\ee

The JT equations of motion of the dilaton field are solved by the sum of the constant term $\Phi_0$ and a varying part $\phi(r)$ which varies according to,
\be 
\phi = c(\beta) \  r.
\label{phi=cr}
\ee

The factor of proportionality $c(\beta)$ in  is temperature dependent and will be fixed in section \ref{sec:dictionary}.  

The proper distance from the horizon will be called $\sigma,$
\be
\sigma=   \int_{\mu}^r dr'  \frac{\mu}{\sqrt{r^2 - \mu^2}}, 
\label{sigma int}
\ee 
from which we find,
\be 
r= \mu \cosh{(\sigma/\mu)}.
\label{r=mu cosh}
\ee
The metric takes the form,
\be 
ds^2 =     -\sinh^2(\sigma/\mu)   dt^2  + d\sigma^2     
\label{met tsig}           .
\ee
Defining $y=\sigma / \mu$ and $t/\mu = i\theta$ the Euclidean continuation has the familiar form of the hyperbolic disc,
\be 
ds^2 = \mu^2  \left[  (\sinh^2{y}) \ \ d\theta^2 + dy^2  \right].
\label{met ytheta}
\ee

\subsection{The Boundary}

\begin{figure}[H]
\begin{center}
\includegraphics[scale=.4]{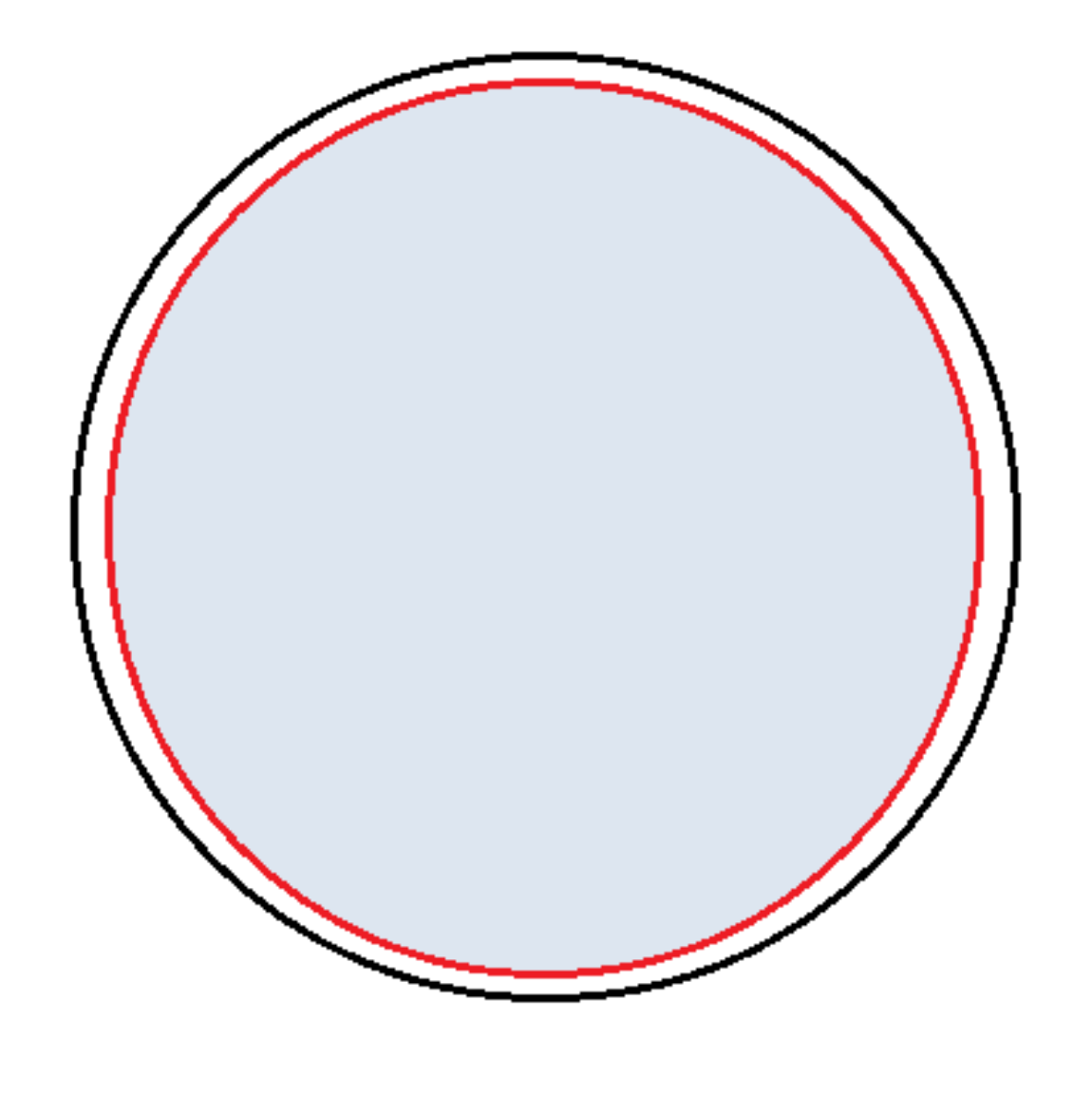}
\caption{Euclidean AdS(2) can be represented by the Poincare disc. The geometry at finite 
temperature is cut off by a boundary shown in red. The bulk geometry is the shaded portion of the disc.}
\label{boundary}
\end{center}
\end{figure}

In the model of \cite{Maldacena:2016upp} the geometry ends at a boundary  located at the value of the radial variable $y$ at which the circumference of the circle $(0<\theta \leq 2\pi)$ is the inverse temperature $\beta,$ 
 $$\sinh{y_b} =  \frac{\beta}{2 \pi \mu}.$$ 
The subscript $b$ stands for boundary. (See figure \ref{boundary})

In terms of $\sigma$ the boundary is at
\be 
\sinh{(\sigma_b/\mu)}= \frac{\beta}{2 \pi \mu}.
\label{shsigb}
\ee
For large $\beta$ (low temperature),
\be 
\sigma_b \approx \mu \log{\lf \frac{\beta}{\pi \mu} \rg}.
\label{sigb}
\ee

The time coordinate $t$ is not the proper time of an observer at the boundary. Following  \cite{Maldacena:2016upp}  we denote the boundary proper time by  $u.$ The relation between $t$ and $u$ is, 
\be 
u                = \lf \frac{\beta}{2\pi \mu}t \rg
\label{def u}
\ee
When considering the relation between JT gravity and the holographic SYK system, the usual convention is that the time conjugate to the SYK Hamiltonian is $u.$

It is convenient when studying the motion of a particle that falls in from the boundary  to introduce a coordinate which measures proper distance from the boundary. Calling it $\rho,$ 
\be 
\rho = (\sigma_b - \sigma).
\label{def rho}
\ee

\begin{figure}[H]
\begin{center}
\includegraphics[scale=.4]{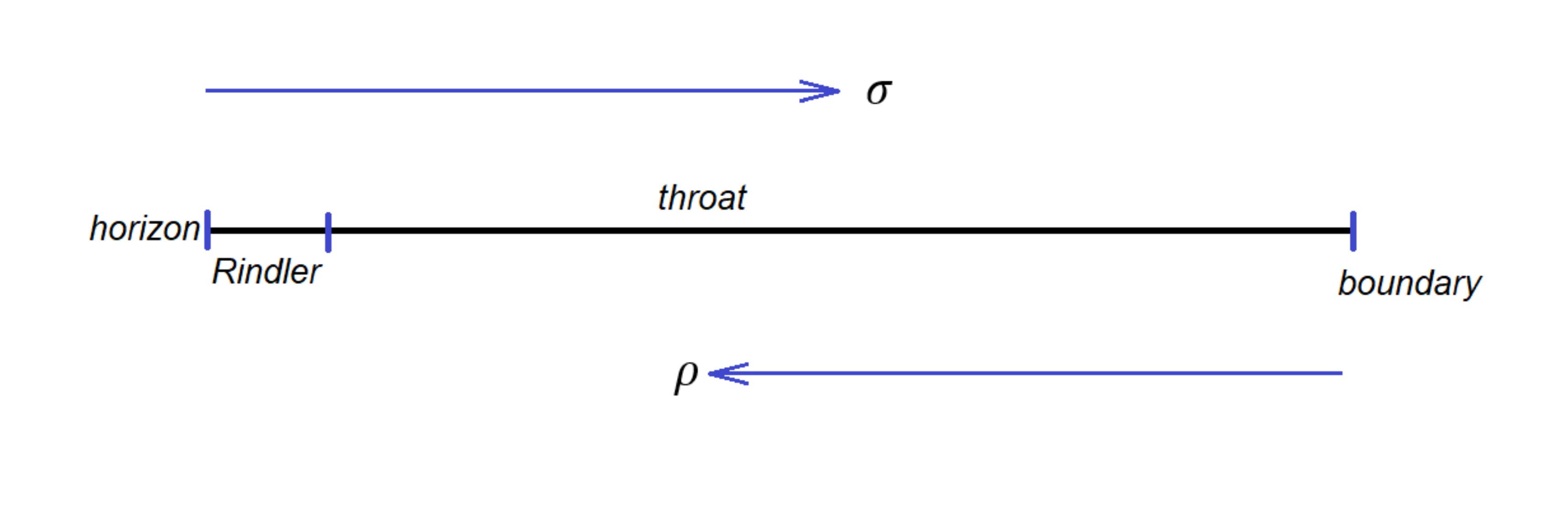}
\caption{The spatial geometry of a near-extremal black hole consists of a long throat whose length diverges in the extremal limit. The throat for non-extremal black holes is bounded from both ends. At the near-horizon end the AdS geometry gives way to a Rindler region and at the far end the throat is terminated at a boundary. The coordinates $\sigma$  and $\rho$ represent proper distance from the horizon and the boundary respectively. }
\label{ads2}
\end{center}
\end{figure}

The $AdS(2)$ spatial geometry and the coordinates $\sigma \ \rho$ are shown in figure \ref{ads2}.

\subsection{Null Trajectory}
A particle dropped from the boundary will very quickly become relativistic and move on an almost light-like trajectory. A light-like trajectory originating at the boundary ($\rho =0$) at time $u=0$ will satisfy,

\be
u =\beta \log{ \lf  \frac{\beta + 2\pi \mu e^{\rho/\mu} }{\beta - 2\pi \mu e^{\rho/\mu}}   \rg } 
- \beta \log{ \lf  \frac{\beta + 2\pi \mu  }{\beta - 2\pi \mu }   \rg } 
\label{null}
\ee
It is easy to show that it will reach the Rindler region in a time $\beta/2\pi.$ Assuming $\beta>>\mu$ then for almost all of that time the trajectory will be in the throat region. 

Equation \ref{null} can be used to convert a function of spatial position $f(\rho)$ to a function of time $f(u)$ along an infalling trajectory. In particular the function $\tb(\rho)$ introduced in 
\cite{Brown:2018kvn}\cite{Susskind:2019ddc} may be thought of as a function of time along the trajectory of an infalling relativistic particle.

\sc
\section{Classical Particle  Motion}
\subsection{Lagrangian, Momentum, and Hamiltonian}

We will begin by deriving the equation of motion of a particle moving in the metric,
\bea 
ds^2 \eq  -f(r) dt^2 +\frac{1}{f(r)}dr^2 \cr  \cr
\eq -f(r) dt^2 + d\sigma^2  
\label{emblack}
\eea
and an abelian gauge field of the form,
 \bea
 A_{\sigma} \eq 0 \cr \cr
 A_t(\sigma) \eq  - (r-r_b)E.
\label{A} 
 \eea
 Here $E$ is constant and represents a uniform electric field and $r_b$ is the value of $r$ at the boundary.

The standard Lagrangian is,
\be  
\CL = -m\sqrt{f(r) -\dot{\rho}^2} +e A_t(\rho)  . 
\label{Lagrangian}
\ee
where dot means derivative with respect to $t.$

The momentum conjugate to $\rho$ is given by,
\bea  
P \eq \frac{\partial \CL}{\partial \dot{\rho}} \cr \cr
\eq 
\frac{\dot{\rho}}{\sqrt{f(r)- \dr^2}},
\label{def P}
\eea
and the Hamiltonian by,
\be 
H_t = P \dr - \CL = \frac{mf}{\sqrt{f(r)- \dr^2}} -eA_t(\rho).
\label{Ht}
\ee
 Note that the  Hamiltonian $H_t$ is conjugate to $t$ and not to the boundary time $u.$ 
 The Hamiltonian conjugate to  boundary time  will simply  be denoted by $H$ without any subscript. The relation between $H$ and $H_t$ is,
\be
 H = H_t  \ \frac{dt}{du } =  H_t   \lf \frac{2\pi\mu}{\beta} \rg.      
 \label{H}    
\ee
  
\subsection{Force}

The radial  force $F_t$ on the particle is defined to be $dP/dt$ and is given by,
\bea 
  F_t \eq  \frac{\partial \CL}{\partial \rho} \cr \cr
  \eq -\frac{m}{2}  \frac{\partial_{\rho}f}{\sqrt{f-\dr^2}}   +e     \frac{\partial A_t}{\partial \rho}\cr \cr
  &=& 
   -\frac{m}{2}  \frac{\partial_{r}f}{\sqrt{f-\dr^2}} \frac{dr}{d\rho}   + e     \frac{\partial A_t}{\partial r}\frac{dr}{d\rho} \cr \cr
   \eq     \frac{m}{2}  \frac{\partial_{r}f}{\sqrt{f-\dr^2}}\sqrt{f}  + e     \frac{\partial A_t}{\partial r} \sqrt{f}
 \label{def Ft1}
\eea
  
From \ref{Ht} we find,
  \be 
  F_t = \frac{\partial_r f}{2\sqrt{f}}  (H_t+eA_t)  - e     \frac{\partial A_t}{\partial r} \sqrt{f}
 \label{def Ft2}
  \ee
  
The force $F_t$ represents the rate of change of   momentum $P$ with respect to $t,$
\be
F_t = \frac{dP}{dt}.
\label{Ft3}
\ee 
The rate of change of $P$ with respect to boundary time will be denoted without a subscript,
\be
F = \frac{dP}{du} = F_t   \lf \frac{2\pi\mu}{\beta} \rg.
\label{F1}
\ee
  Thus one finds,
  \bea
 F \eq  F_G + F_ E \cr \cr
F_G   \eq \frac{\partial_r f}{2\sqrt{f}}  (H+eA_u) \cr  \cr
 F_E \eq - e     \frac{\partial A_u}{\partial r} \sqrt{f},  
\label{F2}
  \eea
  where $$A_u = A_t \frac{dt}{du} = A_t \lf \frac{2\pi \mu}{\beta}  \rg.$$ The subscripts $G$ and $E$ refer to gravitational and electric.
  
  Using  \ref{sigb}, \ref{def u},  \ref{def rho}, and \ref{A}, in the throat region equation \ref{F2} may be written as,

  \bea
  F \eq F_G + F_E,   \cr \cr
  F_G \eq    \frac{H}{\mu}    + eE (e^{-\rho/\mu}-1)    \cr \cr
  F_E \eq eEe^{-\rho/\mu}
    \label{F3}
  \eea
  
The factors of $e^{-\rho/\mu}$  in \ref{F3} go to zero as we move away from the boundary at $\rho=0.$ In particular this means that $F_G$ quickly tends to a constant value 
$$F_G \to  \frac{H}{\mu}    - eE   $$
while $F_E$ rapidly decreases as the particle moves away from the boundary.

  For an electrically neutral particle the force has the simpler form,
  
\be
  F= \frac{\partial_r f}{2\sqrt{f}}  H  = \frac{H}{\mu}.
  \label{Fneut}
\ee

\sc
\section{Relation to SYK }\label{sec:dictionary}

\subsection{Qi-Streicher Formula}
In the Majorana-SYK theory acting with a fermion operator $\psi$ at time $u=0$ creates a particle at rest at the boundary, which subsequently falls into the throat. 
The Heisenberg  operator $\psi(u)$ initially has size $1$. The size grows with time, but the energy is conserved. This is the process described below by the Qi-Streicher formula \ref{QS}.  The energy of such 
 an initial size-$1$ particle is  what we have been calling  $H$ in equations like \ref{Fneut}. 

As we will discuss in section  \ref{sec: operator growth} the momentum-size correspondence   can be expressed in the form  \cite{Lin:2019qwu}\cite{Susskind:2019ddc},
\be 
P = \frac{d\CC}{du}
\label{P=dC/du}
\ee
where $\CC(t)$ is the size of the operator $\psi(t).$

Qi and Streicher \cite{Qi:2018bje} have calculated the size of $\psi(u)$ in \msyk \ to be,
\be 
\CC(u) = 1 + 2\lf \dfrac{\beta \CJ}{\pi}  \sinh{(\pi u/\beta)}   \rg^2.
\label{QS}
\ee

From \ref{P=dC/du} and \ref{QS} we find that in the throat,
\be 
P=4\CJ^2 u,
\label{P=4JJu}
\ee
and the force which is defined as $dP/du$ is given by the constant value,
\be 
F=4\CJ^2.
\label{F=4JJ}
\ee
From \ref{Fneut} and \ref{F=4JJ} we obtain the following expression for $H$,
\be 
H = 4\mu \CJ^2.
\label{H=4muJJ}
\ee

To determine $\mu$ in terms of SYK parameters we turn to black hole thermodynamics  and compare the specific heat of a near-extremal black hole with that of the SYK model.

\subsection{Specific Heat at T=0: Relating $\mu$ and $\CJ$ }\label{sec: specific}

We would like to express $\mu$ in terms of the SYK parameters $N, \CJ, q.$ To do so we 
will equate the zero-temperature specific heat of the JT gravity solution with the corresponding SYK result. We begin with the dilaton which plays the role of area in JT. The dilaton has a constant  background part $\Phi_0$ and a varying part $\phi.$

The dilaton profile $\phi(\sigma)$ is determined by solving the equation of motion and setting the boundary value to a fixed constant of order $\Phi_0.$ The precise value is not important but for convenience  we will choose, 
\be 
\phi_b = \frac{\Phi_0}{\log{2}}.
\label{d6}
\ee

One finds
\be 
\phi(\sigma) = \frac{2   \Phi_0 }{\log{2}}    \frac{\pi \mu}{\beta} \cosh{\sigma/\mu}.
\label{d7}
\ee

The extremal entropy is given by $$S_0 = \frac{\Phi_0}{4G}.$$ For large $q$ it has the value  $\frac{1}{2}N\log{2}$. Thus,
\be 
\frac{\Phi_0}{2G} = N\log{2}.
\label{d8}
\ee

The near-extremal entropy is given by  $1/4G$ times the dilaton at the horizon which
from \ref{d7} is,
\be 
S= \frac{\Phi}{4G}= \frac{\Phi_0}{4G} +   T \frac{\Phi_0 \pi \mu}{2 \log{2} \  G}
\label{d9}
\ee
where $T=1/\beta.$

Using \ref{d8},
\be 
\frac{dS}{dT} = \pi \mu N.
\label{d10}
\ee

To determine $\mu$ we use the SYK value for $dS/dT.$ In the large $q$ limit the SYK value has been calculated analytically  \cite{Maldacena:2016hyu},
\be 
(dS/dT)_{SYK} = \frac{\pi^2 N}{q^2 \CJ}.
\label{d11}
\ee

Equating \ref{d10} and \ref{d11},
\be 
\mu =  \frac{\pi}{q^2 \CJ}
\label{d12}
\ee
for large $q$.

Finally, having obtained $\mu$ in terms of SYK parameters, we may go back to 
 \ref{H=4muJJ} and obtain $H$,
\be 
H=  4\pi  \frac{\CJ}{q^2}.
\label{H(Jq)}
\ee
\subsection{Mass of the Boundary}

We have treated the boundary as if it were frozen at $\sigma = \sigma_b $ with $\sigma_b$ given by \ref{shsigb}. To put it another way, we have treated the boundary as if it has infinite mass.
 In practice that is a good approximation, but in fact the boundary is a dynamical object which can recoil and move when a force is applied to it. For example,
when a particle is dropped in from the boundary the boundary will recoil so that its momentum will be equal and opposite to the momentum of the particle \cite{Susskind:2019ddc}. However, the boundary itself behaves as a very heavy non-relativistic particle whose mass is, 
\be 
M=    \frac{N}{\mu}.
\label{M=N/mu}
\ee

Using \ref{d12} we may write,
\be 
M = \frac{1}{\pi} Nq^2\CJ.
\label{M(qJ)}
\ee
\
The fact that the boundary mass is order $N$ means that the recoil motion is usually negligible.
\bn

\subsection{Summary of Preliminaries}  

For $q>>1$  the relation between the JT and SYK variables that we will need is given by,
\bea 
\mu \eq \frac{\pi}{q^2 \CJ} \ \ \ \ \ \ \ \ \ \ \ (a)   \cr \cr 
\frac{\Phi_0}{2G} \eq N\log{2}    \ \ \ \ \ \ \ \  (b)   \cr \cr
H\eq  4\pi  \frac{\CJ}{q^2}    \ \ \ \ \ \ \ \ \ \ \  (c)    \cr \cr
M \eq      \frac{1}{\pi}    N q^2 \CJ.   \ \ \ \ \ \ \ \  (d)   \cr \cr
 F &=& 4\CJ^2    \ \ \ \ \ \ \ \ \ \ \ \  (e)
\label{d13}
\eea
where in \ref{d13}(c,d) the quantites $H$ and $F$ refer  to the energy and
 the force on a size-1 particle in the throat.

\sc

\sc
\section{Operator Growth in Majorana-SYK} \label{sec: operator growth}

Qi and Streicher   	\cite{Qi:2018bje}  have given a rigorous discussion of operator growth in SYK at low temperature. In this paper we make no pretense of rigor.
We rely on intuitive ``epidemic" arguments  \cite{Susskind:2014jwa} which reproduce the results of 	\cite{Qi:2018bje}  in the throat region and which are easy to generalize to the charged case. 
The argument is best justified in the large $q$ limit in which $1<<\beta \CJ << q^2 << N.$ Later we will discuss the extrapolation to fixed but large $q^2$ and $\beta \CJ >> q^2.$

During the passage through the throat which lasts for a boundary-time $\sim \beta$ the size and complexity of an evolving simple operator are indistinguishable and we will use the same symbol for both; namely $\CC.$

\sc
\subsection{Exponential Growth in Circuit Time}

\begin{figure}[H]
\begin{center}
\includegraphics[scale=.4]{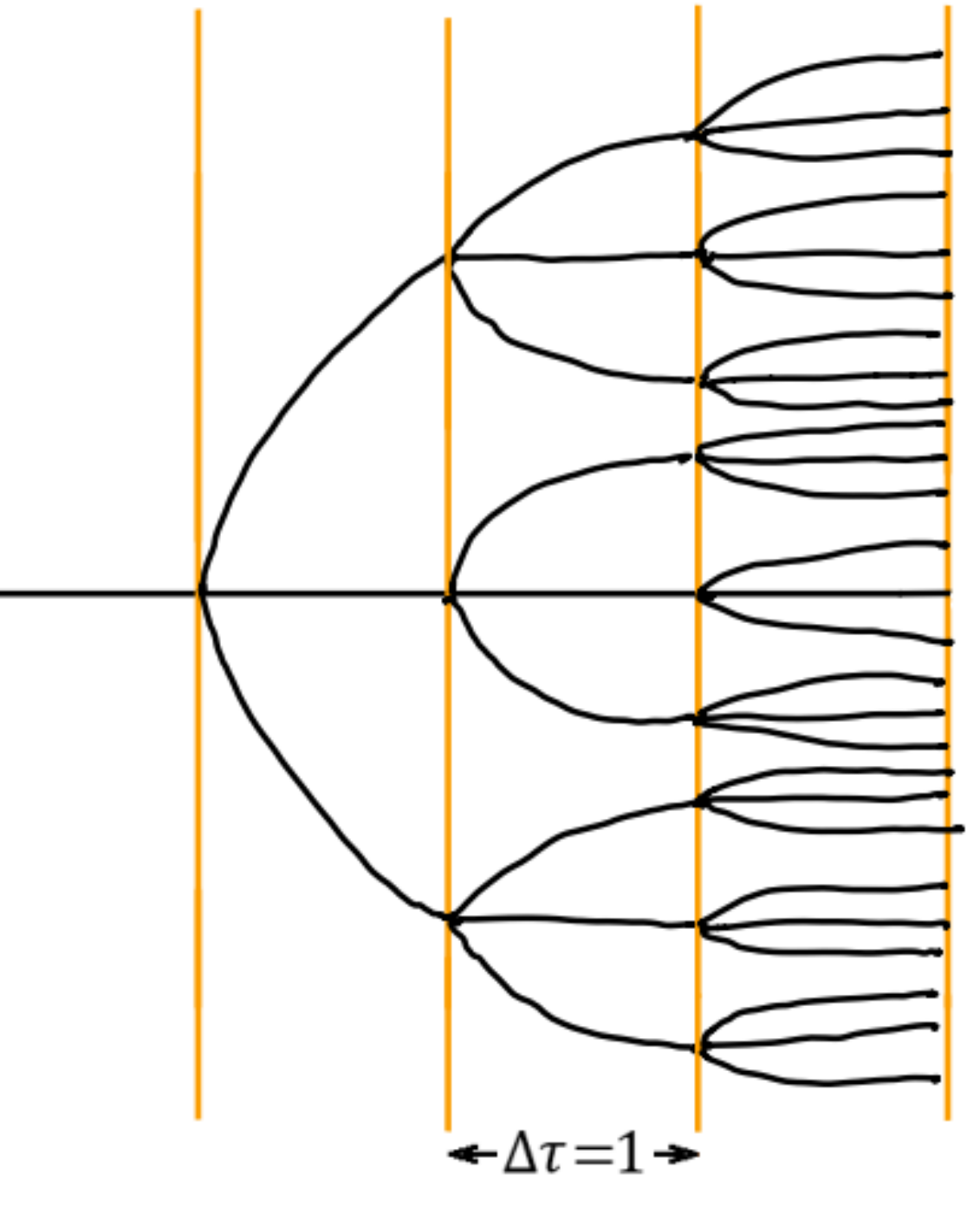}
\caption{Branching diffusion. The diagrams controlling the large $q$ behavior of operator growth in SYK are tree like. They grow exponentially with depth. The number of leaves of the tree is proportional to the number of vertices. The tree-like behavior and exponential growth breaks down at the scrambling time which only happens well after the particle has left  the Throat and passed through the Rindler region.}
\label{branching}
\end{center}
\end{figure}

At large $q$  the evolution of a single fermion operator   is described by diagrams of the form shown in figure \ref{branching}. The diagram represents the ``branching diffusion" pattern shared by SYK and string theory. It is simplified so that it shows an uniform rate of branching with respect to some yet-to-be-determined time variable $\tau.$ This is an simplification  which is only correct in an average sense but it's good enough for our purposes. The time variable $\tau,$ which we will call ``circuit time" is not assumed to be the boundary time $u$ or even linear\footnote{The nonlinear relation between circuit and ordinary time was also necessary in the quantum-epidemic model of Qi and Streicher  in order to reproduce the results of their calculation of growth.} in $u.$

Identifying the number of endpoints (leaves of the tree) as the size $\CC$  of the growing operator, the branching diffusion implies an exponential growth of size in $\tau$. In the $q$-local version of SYK a fermion will split into $(q-1)$ offspring. It is natural to write,
$$\frac{d\CC}{d\tau} = (q-2)\CC. $$ However by redefining $\tau$ we may eliminate  the factor $(q-2)$ and write,
\bea 
\frac{d\CC}{d\tau} \eq \CC  \cr \cr
\CC(\tau) \eq e^{\tau}.
\label{g1}
\eea
We have assumed $\CC(0) =1$ in line with the fact that the initial fermion operator in the upcoming SYK calculation has size $1.$  

 As remarked above, the circuit time $\tau$ is not
the usual asymptotic time $u,$ nor is it linear in $u$. One might think that \ref{g1} has no content other than as a definition of $\tau$ but that's not true. The invariant fact expressed by figure \ref{branching} and equation \ref{g1} is that the number
 of leaves of the tree (the size) is proportional to the number of vertices. By analogy with quantum  circuits the number of vertices represents the growing complexity of $\psi(t).$ This is the basis for the claim that size and complexity are proportional to one another.

\subsection{Why Circuit Time Slows Down}

Let us digress for a moment to explain why the circuit time is not proportional to the time $u$ conjugate to the SYK Hamiltonian. Consider the energy carried by each line in the diagram of figure \ref{branching}. An incoming energy $\omega$ will split at each vertex so that the average energy carried by a line will decrease exponentially with $\tau.$ This naturally implies that the time $\Delta u$ to the next vertex will grow exponentially with $\tau.$ In terms of the epidemic model it means that the rate of contact between participants slows down in real time. This slowdown was noted in   \cite{Susskind:2019ddc} where the time between  vertices $\Delta \tau$ was denoted $\tb.$ It was also commented on in \cite{Qi:2018bje}.

\bn
 
 The proportionality of size and complexity can not go on forever. It and
 and the tree-like nature of the evolution must break down at the scrambling time when the size becomes comparable to the number of fermionic degrees of freedom, $N$. At that point the size saturates but the complexity continues to grow linearly with $u.$ But by this time the particle has long passed through the throat and is in the Rindler region.

We will assume that boundary time $u$ and circuit time $\tau$ are monotonically  related, $$u=u(\tau), \ \ \ du/d\tau > 0.$$ 
Following \cite{Susskind:2019ddc} we define,
\be 
\frac{du}{d\tau} = \tb.
\label{g2}
\ee
The first of equations \ref{g1} takes the form
\be 
\frac{d\CC}{du} \frac{du}{d\tau}  =  \tb \frac{d\CC}{du} =\CC
\label{g3}
\ee

\subsection{Momentum and Size}

In \cite{Susskind:2019ddc} a specific form of the momentum-size correspondence was proposed, 
\be 
\tb P =\CC.
\label{g4}
\ee
Combining \ref{g4} with \ref{g3} implies,
\be 
P=\frac{d\CC}{du}.
\label{g5}
\ee

Equations \ref{g4} and \ref{g5} are two forms of the momentum-size correspondence which in the scaling region, i.e., the throat, are equivalent. Which of them is more fundamental is not obvious but the version \ref{g5} was derived in a rigorous manner
from the  $SL(2R)$ symmetry of AdS(2). Moreover it holds not only through the throat but also through the scrambling process in the Rindler region. 

Note that one may start with \ref{g5} and reverse the argument to derive \ref{g4}.

\subsection{Boundary  Time and Circuit Time}

Let us consider the functional relation $u(\tau).$ One way to obtain it is to use the approximate scale invariance of SYK at low temperature. It implies that $\CC(u)$ should be a power of $u$. Furthermore it should be time-symmetric and smooth at $u=0$. The simplest dimensionally consistent  assumption is 
\be 
\CC(u) = c \CJ^2 u^2
\label{g6}
\ee
with $c$ being a constant.  Qi and Streicher  \cite{Qi:2018bje}  calculate that at large $q$  the size grows according to \ref{QS},
\bea 
\CC(u) &=& 1 + 2 \lf \frac{\beta \CJ }{\pi} \sinh{(\pi u/\beta)} \rg^2  \cr \cr
&\approx& 2\CJ^2 u^2.    \ \ \ \ \ \ \ \ \ \ \ (u<< \beta \ \rm \ in \ throat)
\label{g7}
\eea
From \ref{g7}  we read off that the constant in \ref{g6} is $c=2.$

Next we note that 
$$\tb =  \frac{ \CC}{d\CC/du} $$
 so that 
\be 
\tb = u/2,
\label{g8}
\ee
and from $du/{d\tau} = \tb$
we find,
\be 
u=   {\CJ}^{-1} e^{\tau/2}.
\label{g9}
\ee
where the factor $ {\CJ}^{-1} $ is necessary for dimensional consistency and to match \ref{g7}.

\subsection{Momentum and Force}

The conjectured 
  momentum-size correspondence, if correct, would imply that the momentum  is given by,
\be 
P=d\CC/du =4\CJ^2 u.
\label{g10}
\ee
This may be interpreted in terms of a  constant force $F$ acting on the particle,
\be 
F = dP/du =4 \CJ^2.
\label{g11}
\ee
which agrees with \ref{F=4JJ}.

\subsection{Ladder Graphs}

\begin{figure}[H]
\begin{center}
\includegraphics[scale=.45]{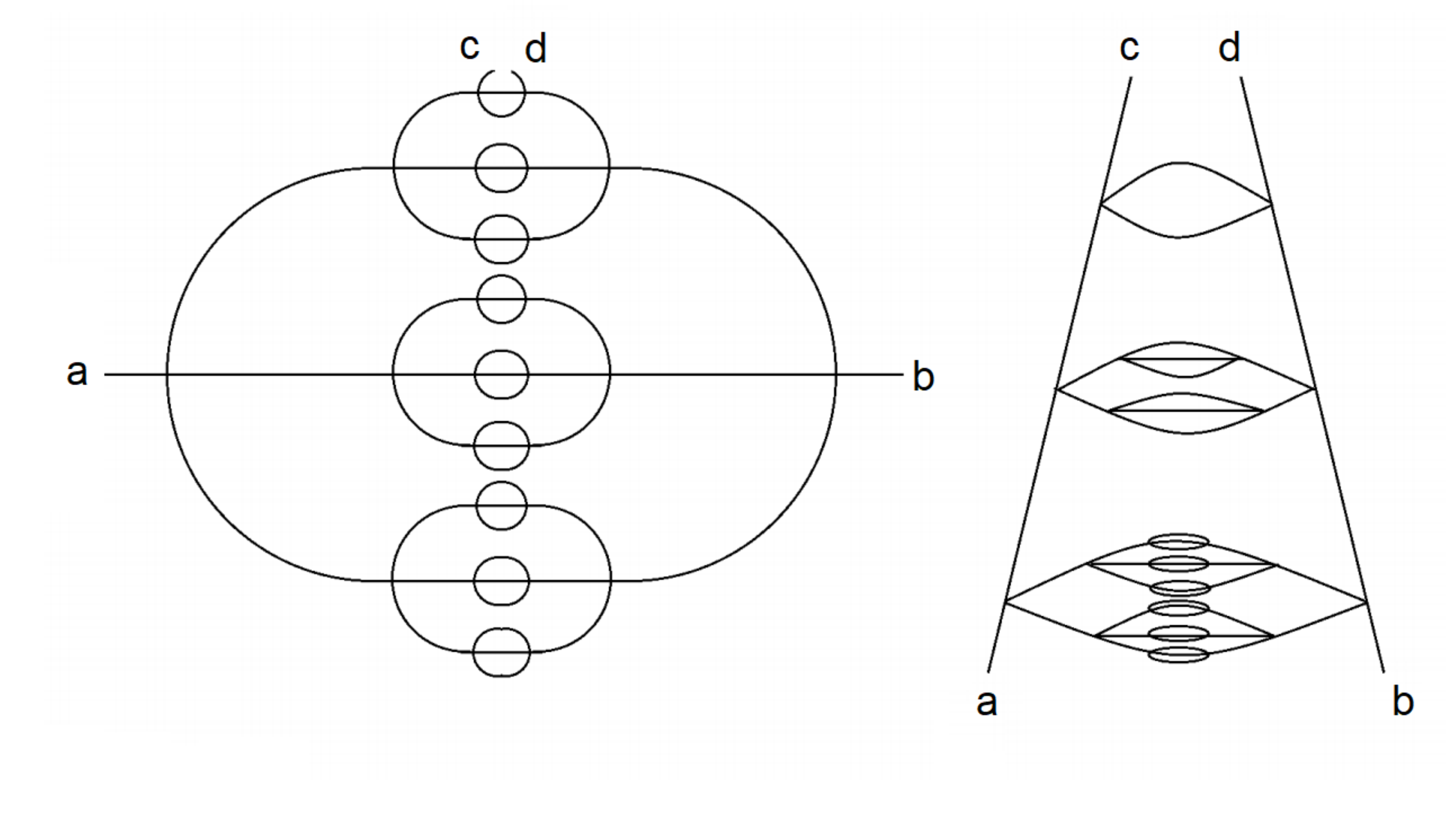}
\caption{Diagrams like figure \ref{branching} represent amplitudes for a given number of fermions, in this case $n=27$. To construct probabilities the amplitude must be squared. In the left diagram the gap in the uppermost line represents the counting of fermions at a fixed time. The ladder diagram on the right is exactly the same diagram as on the left, but redrawn to emphasize the ladder nature of the diagram.}
\label{ladder}
\end{center}
\end{figure}

At this point we will digress  very briefly to comment on the relation between the rate equation    \ref{g1} and the ladder-sum for four-point out-of-time-order-correlators. The diagrams of  figure \ref{branching} represent amplitudes which need to be squared in order to give probabilities. Accounting for the usual SYK disorder averaging the  leading melon diagrams are like the one in the left panel of figure \ref{ladder}. In order to calculate the average size at a given time one needs to count the number of fermionic lines in the intermediate state. This is done by inserting the number operator\footnote{The definition of the number operator that we use was given by Qi and Streicher \cite{Qi:2018bje}. The number operator is defined on two copies of the $N$ fermion system which are identified with the bra and ket sides of the diagrams in figure \ref{branching}. 
}  $[(\psi_i)_L , (\psi_i)_R]$  according to the prescription in  \cite{Qi:2018bje}. Diagrammatically it  is denoted by the open gap on one of the intermediate lines. The open gap can be on any of the intermediate lines and the sum over such diagrams represents the probability for a given size. 

In fact the symmetry of the diagram is such that the sum just gives a multiplicative factor equal to the number of fermion lines.

The relation with four-point ladder diagrams is clear if we redraw the left panel in the equivalent
form in the right panel. The two diagrams are equivalent but the right panel has been drawn
to emphasize the ladder-like nature of the graphs. 

One question that figure \ref{ladder} raises is why the lower rungs of the ladder have 
more internal structure than the upper rungs. The answer is that in the model with uniform
spacing of vertices in  $\tau$ the circuit time between the beginning and end of a rung is greatest 
at the bottom of the ladder and grows smaller as we proceed up the ladder. The actual ladder kernal is a sum of diagrams, but which diagram dominates depends on the time-interval between where it connects with the rails.

From the ladder diagram point of view the exponential growth of the size is characteristic of 
the graviton Regge trajectory.

\sc
\section{Charged Particles in U(1)-invariant SYK} \label{charge}

Now we turn to the main subject of the paper---operator growth and its relation to electric forces in the
 $U(1)$-invariant version of the SYK model. The model has 	a global ungauged conserved charge. Positive and negative charged fermions  are coupled in a $q$-local manner. The $U(1)$ symmetry requires that 
 each term in the Hamiltonian has $q/2$ fermions of each sign. Schematically,
 \be 
 H= \sum j_{....} \psi_1^{\dag}...\psi_{\frac{q}{2}}^{\dag}  \psi_1...\psi_{\frac{q}{2}}
 \label{c1}
 \ee
 with\footnote{The definition of $\CJ$ is the same as in Majorana-SYK except that the combinitorial factor $q!$ in M-SYK is replaced by $(q/2)!^2$ in $U(1)SYK.$}
 \be 
 \la j^2\ra =       \frac{2^{q-1} \CJ^2 }{q^2 N^{q-1}}  \lf q/2   \rg !^2
  \label{c2}
 \ee
 and $\CJ$ is to be held fixed as $N$ and $q$ are varied. 
 
 The model we will use for studying operator growth is a simple variant of the perturbative epidemic model  \cite{Susskind:2014jwa}
  which is justified in the limit $(1<< \beta \CJ << q^2 <<N.)$ Later we will discuss its extrapolation to $(1  << q^2<< \beta \CJ <<N.)$

\subsection{The Chemical Potential}\label{sec: The Chemical Potential}
 
\begin{figure}[H]
\begin{center}
\includegraphics[scale=.4]{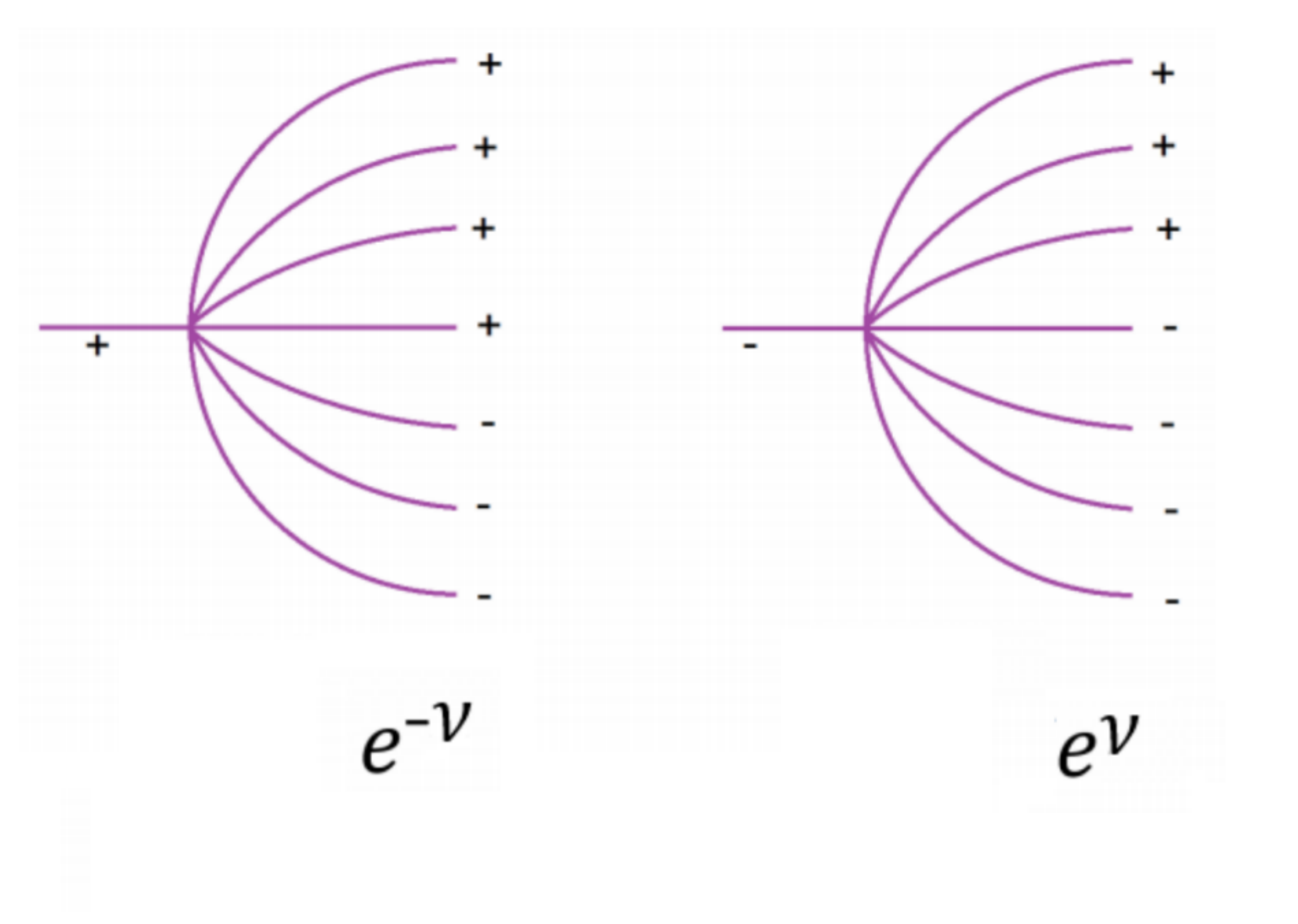}
\caption{$q=8$ ``pitchfork diagrams" representing the basic unit of operator growth in the U(1)-SYK model.The expressions under each vertex indicate its weighting. } 
\label{pitchforks}
\end{center}
\end{figure}

 The vertices governing the growth of size are shown in figure \ref{pitchforks}.
\bn
The diagrams in figure \ref{pitchforks} are to be read from left to right,  the propagators  being time-ordered.

To vary the charge of the black hole a chemical potential $\nu$ is 
introduced\footnote{In our convention the chemical potential is dimensionless in contrast to the usual convention in which it has units of energy. } 
into the thermal density matrix as an extra factor $e^{\nu (n_+ - n_-)},$ where $(n_+-n_-)$
is the total fermion charge.
It's effect is to modify the bare propagators connecting the vertices. A line labeled  $\pm$ carries a factor $e^{\mp \nu}$ as in figure \ref{pitchforks} 

Let us see how these factors comes about. The bare propagators are defined in the non-interacting theory ($\CJ =0$) in which the different fermionic coordinates do not interact. Therefore it is sufficient to concentrate on a single complex fermionic degree of freedom. We may think of the fermionic system in terms of Pauli matrices,

\bn

\bn

\bn

\bn

\bea
\psi^{\dag} &=& \sigma_+  = (\sigma_x+i\sigma_y)/2 \cr \cr
\psi &=& \sigma_-  = (\sigma_x-i\sigma_y)/2 \cr \cr
n &=& \sigma_z /2 
\label{c3}
\eea

The bare propagator is defined in terms of the non-interacting two-point function,
\be 
\la \psi^{\dag}  \psi \ra = \frac{1}{2} \ \rm Tr \it \  \rho \psi^{\dag}  \psi
\label{c4}
\ee
where the density matrix is 
\be 
\rho = e^{\nu n} = e^{\nu \sigma_z/2}
\label{c5}
\ee
One finds,
\bea
\la \psi^{\dag}  \psi \ra \eq \frac{1}{2} e^{\nu/2}  \cr \cr
\la \psi  \psi^{\dag} \ra \eq \frac{1}{2} e^{-\nu/2}
\label{c6}
\eea

\bea
\la \psi  \psi^{\dag} \ra \eq \frac{1}{2} e^{-\nu/2} \cr \cr
\la \psi^{\dag}  \psi \ra \eq \frac{1}{2} e^{\nu/2}  
\label{c6}
\eea
Thus a line representing a positive fermion carries a factor $ \frac{e^{\nu/2}}{2}$ and a line representing a negative fermion carries a factor $ \frac{e^{-\nu/2}}{2}$.

\begin{figure}[H]
\begin{center}
\includegraphics[scale=.45]{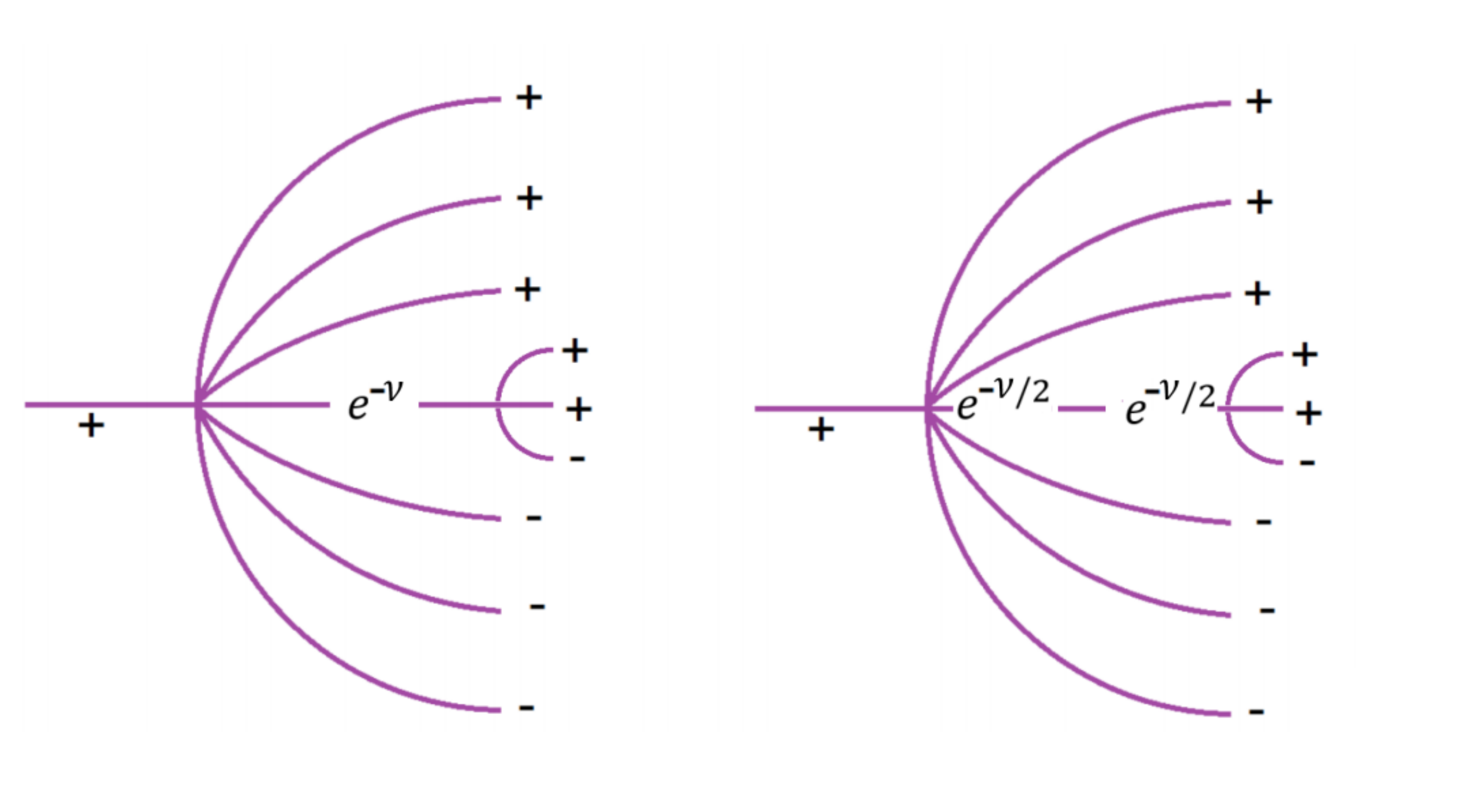}
\caption{Splitting and redistributing $e^{\nu}$. A propagator representing a positive fermion is shown weighted by a factor $e^{-\nu}$ in the left panel. In the right panel that same factor is redistributed to the vertices connected by the propagator.  }
\label{splitmu}
\end{center}
\end{figure}

 It is convenient to redistribute these factors by splitting each $e^{\pm \nu/2}$ into two factors 
$e^{\pm \nu/4}e^{\pm \nu/4}$ and bringing them to the vertices as in figure \ref{splitmu}. when this is done each vertex with an incoming $\pm$ line will be weighted with a factor $e^{\pm\nu/2}.$ In addition the entire graph will have an overall factor $e^{\nu/2}$ which does not depend on the graph.  The coupling $\CJ$ governing the strength of each vertex is thereby ``renormalized " from $\CJ$ to $e^{\pm \nu/2} \CJ.$ This is indicated in figure \ref{pitchforks}.

The factors  $e^{\pm \nu/2}$ occur in amplitudes---roughly speaking, the amplitude that the initial fermionic perturbation evolves to a given number of positive and negative constituents---but  the  rate equations in the next section are about probabilities. This means that in the weighting of each graph the factors  $e^{\pm \nu/2}$ must be squared to give a weight
$e^{\pm \nu}$ to each vertex.

\subsection{The Rate Equations}\label{sec: The Rate Equations}

Equation \ref{g1} is the simplest possible rate equation that one can write down. In the case of $U(1)$-SYK the equations are similar but  more complicated since one has to keep track of two species of fermions---those with positive charge and those with negative charge.

Define the partial sizes $n_{\pm}$ to be the average number of $\pm$ charged fermions in an operator. The total size  $(n_+ +n_-)$ will again be denoted $\CC.$ If the process is initiated by a $\pm$ charged fermion operator,  then the size after circuit time $\tau$ will be denoted by $\CC(\tau)_{\pm}$. Specifically the size of $\psi^{\dag}(\tau)$ is $\CC(\tau)_{+}$ and similarly the size of  $\psi(\tau)$ is $\CC(\tau)_{-}$.

 The rate equations for the tree-like graphs that dominate the large $q$ model should have several properties:
 
 \begin{enumerate}
 \item They should be linear. Once a splitting has occurred the different branches do not interact.
 \item For reasons explained above, the splitting of a $\pm$ fermion should be weighted by a probability  $e^{\mp \nu.}$ This is indicated in figure \ref{pitchforks}.
 \item The equations should conserve the value of $(n_+ - n_-).$ This is to insure charge conservation.
\end{enumerate}  

 The most general rate  equation with these properties is,

\be
\frac{d}{d\tau}\begin{bmatrix} 
n_+   \\
n_-  \\
\end{bmatrix}
\quad
=
\begin{bmatrix} 
e^{-\nu}\    &   e^{\nu}        \\
e^{-\nu}\    &   e^{\nu}       \\
\end{bmatrix}
\quad
\begin{bmatrix} 
n_+   \\
n_-  \\
\end{bmatrix}
\quad
\label{c7}
\ee

The eigenvalues and eigenvectors\footnote{The ket-like notation here does not indicate a quantum state, but only the real column vector whose entries are $n_+, n_-.$ } of the matrix in \ref{c7} are,
\bea
\lambda_1 \eq 0 \cr \cr
|\lambda_1) \eq   
\begin{bmatrix} 
e^{\nu}   \\
-e^{-\nu}  \\    
\end{bmatrix}
\quad
 \cr  \cr
 \lambda_2 &=& 2 \cosh{\nu} \cr \cr
 |\lambda_2) &=& 
\bm
1  \\
1   \\
\em
\quad
\label{c8}
\eea

We will be interested in solutions of the rate equations in which the initial state is either one positive or one negative fermion. We denote these initial states by $|\psi(0))_{\pm}.$ By definition,
\bea 
|\psi(0))_+&=&
\bm
1 \\
0  \\
\em
\quad  \cr \cr
|\psi(0))_-&=&
\bm
0 \\
1  \\
\em
\quad  
\label{c9}
\eea
The solution of the rate equations with these initial conditions is,
\be 
|\psi(\tau))_{\pm}= \frac{1}{2\cosh{\nu}}
\bm
e^{\tau {\mp} \nu} + e^{\pm \nu} \\
e^{\tau {\mp} \nu} - e^{\mp\nu}  \\
\em
=
\bm
n_+ \\
n_-  \\
\em_{\pm}
\label{c10}
\ee
and finally,
\be 
\CC_\P = n_+ + n_- = \frac{1}{2\cosh{\nu}} (2e^{\tau \mp \nu} +e^{\pm \nu}-e^{\mp \nu} )
\label{c11}.
\ee

In general when we compare with the GR calculation the presence of a chemical potential will back-react and influence the metric, but by working to lowest order in 
 in $\nu$  back-reaction  can be ignored. Thus from now on we will work to linear order in 
 $\nu,$
\be 
\CC_\P =  e^{\tau} (1 \M\nu)  \P \nu 
\label{c12}
\ee
This is the total size for an initial state with one $\P$ fermion.

Using \ref{g9} we find the sizes as a function of $u$,
\be
\CC_{\pm}(u) \ap  (1\mp\nu)\CJ^2 u^2  \pm \nu.
\label{c13}
\ee

For $\CJ u>1$ the second term may be ignored compared to the first,
\be 
\CC_{\pm}(u) \approx  (1\mp\nu)\CJ^2 u^2  \ \ \ \ \ \ \ \ \ \ \  (\CJ u > 1)
\label{c14}
\ee

As an example consider the size of a positively charged operator $\CC_+$ assuming the chemical potential is positive. From \ref{c14} we see that the growth is slowed relative to the case with $\nu=0.$ Since with our convention a positive chemical potential will induce a positive charge, the result implies that like-sign charges retard the operator growth. Similarly opposite charge increases the growth rate. This of course is the holographic equivalent of like-signs repel, unlike signs attract.

\subsection{Relation to OTOCS and Regge Behavior}

The time dependence  of size is rigorously quantified by the evolution of the 4-point out-of-order-correlator. In SYK these OTOCS are dominated by ladder diagrams which are evaluated by iterating a ladder kernel. The ladders  are described by Regge behavior with Regge intercepts given in terms of the eigenvalues of the kernel.  

The matrix in the 
 rate equations \ref{c7} is a simplified version of this kernel. In the limit $\nu \to 0$ the two eigenvalues in 
 \ref{c8} are $\lambda_1=0$ and $\lambda_2=1.$ The second is already present in Majorana-SYK and corresponds to the graviton trajectory. The first is a new feature of $U(1)$-SYK and corresponds to the photon trajectory.

\subsection{The Evolution of Size for Charged Operators}

In the large $q^2,$ fixed $\beta\CJ$ limit the charge of the system and the chemical potential are related by, 
\be
Q=\frac{1}{4}N\nu.
\label{c15}
\ee
Equation \ref{c15} together with
\be 
M= q^2\CJ N
\label{c16}
\ee
imply that equation \ref{c14} can be re-written
 in terms of the black hole mass and the integer-valued $U(1)$ generator $Q.$ 
 \be 
\CC_{\pm} \ap \CJ^2 u^2  \lf   \frac{M}{\CJ Nq^2} \pm \frac{4Q}{N}      \rg
\label{c17}
\ee

Concerning equation \ref{c15}, it  is true in the weak coupling limit and therefore in the limit $q^2>>\beta \CJ.$ It may be derived in the free fermion limit. To see that consider the partition function for a single free fermion mode,
\bea 
Z &=& \rm Tr \it e^{\nu n} \cr \cr
\eq \rm Tr \it e^{\nu \sigma_z/2} \cr \cr
\eq 2 \cosh{\nu/2}
\label{c18}
\eea
The charge $Q$ to first order in $\nu$ is given by,
\bea 
Q\eq \frac{\partial \log{Z}}{\partial \nu} \cr \cr
\eq \frac{1}{4} \nu
\label{c19}
\eea
Extending this to $N$ fermionic modes gives \ref{c15}.

Equation \ref{c15}  is correct only correct for $q^2 >> \beta J$ and must be significantly modified in the more interesting limit of fixed $q$ and very large $\beta \CJ.$ We will discuss this in section \ref{Extrap}.

By using the momentum-size correspondence we will compare \ref{c17} with a general relativity calculation of the motion of charges in a NERN background. For the purpose of that comparison we note that the momentum-size correspondence applied to \ref{c17}  would give,
\be 
P =  \frac{d\CC}{du} \approx \CJ^2 u  \lf   \frac{M}{\CJ Nq^2} \pm \frac{4Q}{N}      \rg
\label{c20}
\ee

\subsection{Note on Charges}

In equation \ref{c15} $Q$ is not exactly the electric charge. $Q$  is the difference 
$(n_+-n_-)$ which we do expect to be proportional to the electric charge, but with a factor of proportionality yet to be determined. Let us call the electric charge $\tilde{Q}.$
It differs from the  integer-valued $U(1)$ generator by a factor $|e|$ representing the fundamental unit of charge,
\be 
\tilde{Q} = |e| Q.
\label{c21}
\ee 

\sc
\section{Motion of Charged Particles}
Returning  to equations \ref{F3}, the force on a charged particle in the throat is given by the sum of two terms: a gravitational force $F_G$ and an electric force $F_E.$
  \bea 
   F_G \eq    \frac{H}{\mu}    + eE (e^{-\rho/\mu}-1)    \cr \cr
  F_E \eq eEe^{-\rho/\mu} 
  \label{m1}
  \eea
  As the particle moves away from the boundary (at $\rho =0$) the electric term quickly becomes negligible, so throughout most of the throat the force is constant and equal to,
  \be 
     F =    \frac{H}{\mu}    - eE 
       \label{m2}
  \ee

The Energy $H$ is conserved and may be evaluated at the start of the infall when the particle is at rest at the black hole boundary.   From \ref{d13}(c),
\be 
H = 4\pi \frac{\CJ}{q^2}.
  \label{m3}
\ee

The electric field $E$ is related to the electric charge of the black hole which we denote $\tilde{Q},$
\be 
E=\tilde{Q}   
  \label{m4}
\ee
The final form for the force on the charged particle in the throat is,
\be 
F =4\CJ^2 \pm \tilde{Q}e
 \label{m5}
\ee

It follows that the momentum of an infalling charged fermion is given by,
\be  
P(u) = u  \lf 4\CJ^2 \pm \tilde{Q}e\rg
  \label{m6}
\ee

\subsection{A Puzzle }

There is something puzzling about equations \ref{def Ft1} and \ref{def Ft2}  (it puzzled me); namely the presence of the term 
$$ \frac{\partial_r f}{2\sqrt{f}} e A_u $$
 in the expression for the gravitational force. Being proportional to the  charge it would seem to be part of the electric force, but that would be very odd since the usual expression for 
 electric force involves only the spatial derivative of $A_u.$  
In fact  the term,
 $$ e\frac{\partial A_u}{\partial \rho } =    
 - e     \frac{\partial A_u}{\partial r} \sqrt{f}$$ 
was identified  as the electric force in \ref{F2}. It   is also the  term we dropped from \ref{F3}  because it  quickly becomes negligible as the particle moves away from the boundary.  However,  it would be a mistake to ignore it altogether.

The resolution of the puzzle is the following: The particle initially experiences an electric force
$-e\partial_{\rho}A_u$  as an  impulse as it starts out at $\rho=0.$ 
The impulse, although short-lived, affects the momentum and  kinetic energy of the particle. By the time the particle has moved a distance $\sim \mu,$ depending on the relative sign of the particle and black hole,  the electric impulse  will have  resulted in the particle having greater or lesser energy than a corresponding neutral particle. 
After the initial kick 
 kinetic energy will be
$H+eA_u.$  

On the other hand the gravitational force on the particle  is proportional to its kinetic energy, so depending on the sign of the charge, once the particle gets away from the boundary the gravitational force will either be less than or greater than that of the corresponding neutral particle.  We see that the origin of the term proportional to
$(H +eA_u)   $ is gravitational.

That the electric force  $-e\partial_{\rho}A_u,$                                                                                                                                                                                                                                                                                                                                                                                                                                                                                                                                                                                                                                                                                         decreases quickly compared to the gravitational force is expected from Regge theory\footnote{From the Regge-pole point of view, gravity corresponds to a trajectory with intercept $2$ while electromagnetism corresponds to a trajectory with intercept $1.$ In the current context this implies a dominance of gravitational force over electric force as the particle gains energy in the throat. }.

\subsection{Comparing SYK and GR} \label{sec: Comparing Results}

Now we come to the main point: the comparison of the forces and acceleration of charged particles in the throat, and the growth of operator size in $U(1)$-SYK.
From the general relativity calculation  \ref{m5} we conclude that the momentum $P$ of an infalling charge is given by \ref{m6},
$$
P =u\lf4\CJ^2 \pm \tilde{Q}e\rg.
$$
On the other hand let us 
  return to the rate equations in section \ref{charge}. The result was   equation \ref{c17} for the growth of size, and (assuming the momentum-size correspondence)  \ref{c20} for the growth of momentum. It is evident that \ref{m6} and \ref{c20} have similar form.
 Equating them  we find,
\be 
e^2 =\frac{4\CJ^2}{N}.
\label{r1}
\ee

Equation \ref{r1} is new information  that fixes the fundamental unit electric charge, in other words the charge of a fundamental $U(1)$-SYK fermion. The fact that it is order $1/N$ means that electric forces are  $1/N$ effects which may seem surprising. 
  In the next section we will see that this has a simple explanation.

\subsection{SYK and KK} \label{sec: SYK}

While \ref{r1} may seem surprising it is in fact entirely natural. What it suggests is that the $U(1)$-SYK system is a realization of Kaluza-Klein theory with a compactification scale which we might have guessed. 

Let us add an additional compact KK direction so that the geometry becomes $AdS(2) \times U(1),$ with the compact circle having radius $R_k.$ The unit of electric charge in such a theory would be given by,
\be 
e^2   = \frac{G}{\Phi_0 R_k^2}
\label{r2}
\ee

From \ref{d8} we see that 
\be 
e^2 \sim \frac{1}{NR_k^2}.
\label{r3}
\ee
Comparing this with \ref{r1} we find,
\be 
R_k \sim 1/\CJ.
\label{r4}
\ee

The scaling of $R_k$ with $1/\CJ$ is hardly surprising since $\CJ^{-1} $ is the only length scale that appears in SYK. In principle $R_k$ might  have had some uncancelled  $N$   dependence, but that did not happen.
The lack of  $q$ dependence in $R_k$ is also interesting.  We may compare \ref{r4}  with  \ref{d12},
\be
\frac{R_k}{\mu}\sim q^2.
\label{r5}
\ee
We see  that $R_k$ grows relative to the $AdS(2)$ length scale as $q$ increases. The significance of this is not entirely clear and one
 might wonder if it is an artifact of weak coupling, not to be trusted for fixed $q$ and
large $\beta \CJ.$  In the next section\footnote{The argument of section \ref{BPF} is due to  Henry Lin and Geoff Penington.  } evidence will be given that  \ref{r5} is robust in the
   strong coupling limit $1<<q^2 <<\beta \CJ.$

\subsection{Boundary Particle Formulation}\label{BPF}

The arguments of the previous sections were derived from the weak coupling limit in which $\beta \CJ$ is held fixed and $q^2\to \infty.$ Nevertheless I believe the results---in particular
the scaling of the KK radius \ref{r4}---may be extrapolated to the strong coupling limit in which $q^2$ is held fixed and $\beta \CJ \to \infty.$

A method which is valid at strong coupling is the boundary particle description \cite{Maldacena:2016upp}. In this section we will use the boundary particle description to compute $R_k$ and see that it is indeed given by \ref{r4}. 

In the Majorana case the boundary particle has a single degree of freedom, i.e.,  its proper distance from the horizon. It moves nonrelativistically with a large mass,  $M=N \CJ q^2$ (see \ref{d13}-d).  In $U(1)$-SYK
the particle has a second $U(1)$ angular degree of freedom. The radial direction is frozen by a potential but the angular direction is a zero-mode. The quantized momentum conjugate to the angular coordinate is the integer-valued charge $Q.$ 

Because of the large mass of the boundary particle the Hamiltonian for the angular motion is non-relativistic,

\be 
H=p^2/2M = n^2/2R_k^2M.
\label{r6}
\ee

We consider the thermal ensemble,
\be 
\rho = e^{-\beta H -\nu n} = e^{-\frac{\beta n^2}{2MR_k^2} - \nu n}                                                                                                                                                                                                                                                 
\label{r7}
\ee
Completing the square and integrating over $n$ gives a partition function,
\be 
Z= e^{\frac{\nu^2 R_k^2 M}{2\beta}}.
\label{r8}
\ee
The charge $Q$ is given by,
\bea
Q &=& \frac{\partial{\log{Z}}}{\partial \nu} =\mu \frac{R_k^2 M}{2\beta} \cr \cr
\eq \nu Nq^2 R_k^2\CJ /\beta
\label{r9}
\eea

In the weak coupling limit the charge and chemical potential are related by \ref{c15} but this relation breaks down badly in the strong coupling limit. For 
$\beta \CJ >> q^2$ the correct relation 
  \cite{Davison:2016ngz} is,
\be 
Q=\frac{\nu N q^2}{16\beta \CJ}
\label{r10}
\ee
Equating these expressions for $Q$ gives,
$$ 
R_k^2 = \frac{Nq^2}{8\CJ M}
$$
and finally using \ref{d13}(d) we find the $q$ and $N$ dependences cancel, 
\be 
R_k = \sqrt{\frac{\pi }{8}} \  \frac{1}{\CJ} \sim \frac{1}{\CJ}
\label{11}
\ee
in agreement with \ref{r4}.  

On the other hand the relation $\mu = \pi /(q^2 \CJ)$   (equation \ref{d12}   in section \ref{sec: specific})
made no use of weak coupling. Therefore the mismatch between the scaling of $R_k$ and $\mu$ is not a weak coupling artifact.

To reiterate, the scaling of the KK radius with either $1/\CJ$ was derived in two ways: the weak coupling method featuring rate equations, applicable for $q^2 >> \beta \CJ$; and the strong coupling boundary particle method. This suggests that at least for some quantities it is possible to calculate by first fixing $\beta \CJ$ and going to large $q^2,$ followed by an extrapolation to large but fixed $q$ and much larger $\beta \CJ.$

\section{Extrapolation to Strong Coupling}\label{Extrap}

In this section we will speculate on why the weak coupling analysis used in most of this paper gives results which are applicable for the strongly coupled limit 
$1<<q^2 <<\beta \CJ.$

Let us consider the consequences of a hypothetical  ``renormalization" of the chemical potential $\nu$ in which it is replaced by a renormalized value $\bar \nu,$
\be 
{\bar{\nu}} = f(q, \beta \CJ)\nu
\ee
with $f$ being an arbitrary function of the dimensionless  SYK parameters $q$ and $\beta \CJ.$  In extrapolating from weak to strong coupling there is no reason why $f(q,\beta \CJ)$
should not change a great deal.

There are two places,
 where the chemical potential enters into our analysis. The first was in the rate equations \ref{c7}  and the second was in equation relating chemical potential and charge \ref{c15}. The thing to note is that if $\nu$ were to be replaced by  $\nb$ in both  places:
\be
 \frac{d}{d\tau}\begin{bmatrix} 
n_+   \\
n_-  \\
\end{bmatrix}
\quad
=
\begin{bmatrix} 
e^{\nb}\    &   e^{-\nb}        \\
e^{\nb}\    &   e^{-\nb}       \\
\end{bmatrix}
\quad
\begin{bmatrix} 
n_+   \\
n_-  \\
\end{bmatrix}
\quad
\label{nubar1}
\ee
 \be 
Q=\frac{1}{4}N\nb 
\label{Qsim nb}
 \ee
 the final outcome would be unchanged.
 In particular the relations  \ref{r1} and the relation between the Kaluza-Klein radius and the SYK energy scale $\CJ$  \ref{r4} would be unchanged. 
 
 From 
 \ref{r10} we see that the function $f$ should be given by
\be
f(q, \beta \CJ) =\frac{q^2}{4\beta \CJ}.
\ee
in the strong coupling limit. The question is whether the same renormalization of $\nu$ would correctly describe the strongly coupled rate equations in \ref{nubar1}.

 Let us first consider \ref{Qsim nb}.  The total charge is given by the equal-time two-point function,
 \be 
 Q =\Tr \  i\rho \  [\psi^{\dag}_i, \psi_i] = i\la \psi^{\dag}_i  \psi_i -hc \ra
 \ee
 with $$\rho = e^{-\beta H +\nu n }.$$

 To first order in the chemical potential the total charge is proportional to
$$\frac{d\la \psi^{\dag}_i\psi_i -hc\ra}{d\nu}$$ in other words,
\be 
\nb \sim \frac{d\la \psi^{\dag}_i\psi_i -hc\ra}{d\nu}|_{u=0}
\ee
 
Now consider the $\nu$ dependence of the rate equation,  which originated from  the dependence of the propagators in figure \ref{splitmu} on $\nu.$ In the weakly coupled limit those propagators had the form given in \ref{c6} but they will be corrected by interactions.
Assuming that the diagrams of importance in the coupled theory are the same as those in the weak limit, but with corrected propagators, then to 
 first order in $\nu$ the rate equations in the interacting theory depend on 
$$\nb \sim \frac{d\la \psi^{\dag}_i\psi_i -hc\ra}{d\nu}.$$ The difference is that in the case of the rate equations the effective value of $\nb$ depends on the two-point functions over a  range of time, not just at $u=0.$ It does however seem reasonable that the renormalized values of $\nb$ for the two purposes are approximately the same. 

\section{Conclusion}
This paper studied the emergence of bulk electric forces in a holographic theory with a global symmetry ---namely the $U(1)$-SYK model. There  main ingredients were:

\begin{enumerate}
\item The momentum-size correspondence that relates the momentum of an infalling particle to the size of the operator that created the particle. 
\item On the SYK side, a simple generalization of the epidemic model for operator growth that described the evolution of size for charged operators in a background of non-zero global $U(1)$ charge.
\item A general relativity calculation of the time dependence of momentum of an infalling charged particle in two-dimensional dilaton-Maxwell gravity.

\end{enumerate} 
The agreement of the SYK and GR calculations requires a specific normalization for unit of electric charge carried by the basic fermion operators $\psi.$ That normalization agrees with  a Kaluza-Klein mechanism in which the KK compactification radius is equal to the inverse energy scale of the SYK theory as in equation  \ref{r4}.

The epidemic model was based on weak coupling arguments and does not obviously apply in the limit $\beta\CJ >> q^2$ where the coupling is strong. Two arguments were given that the weak coupling results could be extrapolated to strong coupling. The first was that by using the strong coupling boundary particle formulation one finds the same result for the KK scale, namely \ref{r4}.

The second argument was that the main effect of strong coupling is to ``renormalize" the chemical potential. If the renormalization is the same in the two places where the chemical potential occurs it will cancel out and reproduce \ref{r4} when  $\beta\CJ >> q^2.$

The charged  epidemic model of section \ref{charge} is crude and a more technically precise version of it would be helpful, especially in clarifying the extrapolation to strong coupling.

One last point seems worthy of mention. The gravitational effect on an infalling particle is truly a bulk phenomenon in that the force is uniform throughout the throat. By contrast, although the electric field is uniform, equation \ref{F3} shows that the electric force on a charged particle is localized within a distance $\mu$ from  the boundary.
The difference is due to the different energy dependence of the coupling to gravity and  electromagnetism  

Nevertheless, despite its localization near the boundary, the electric force  leaves an imprint on the kinetic energy  and therefore  affects the subsequent gravitational force on the particle throughout the bulk.

\section*{Acknowledgements}

I am grateful to Adam Brown, Henry Lin, and Geoff Penington for discussions during the early phases of this work. The boundary-particle formulation of section \ref{BPF} is due to Henry and Geoff.

\end{document}